\documentclass[prb,superscriptaddress,amsfonts,amssymb,amsmath,floats,twocolumn,aps,footinbib,shownopacs]{revtex4-2}
\usepackage[pdftex]{graphicx}
\usepackage{subfigure}
\usepackage{dsfont}
\usepackage{lipsum}
\usepackage{dcolumn}
\usepackage{bm}
\usepackage{color}
\usepackage{physics}
\usepackage{multirow}
\usepackage[pdftex,colorlinks=true, linkcolor=blue,citecolor=blue,filecolor=blue]{hyperref}
\usepackage{comment}
\begin{document}
\title{Thermalization Fronts in the Hubbard-Holstein Model} 

\author{Antonio Picano} 
\affiliation{JEIP, UAR 3573, CNRS, Coll\`ege de France, PSL Research University, 11 Place Marcelin Berthelot, 75321 Paris Cedex 5, France}
\author{Marco Schir\`o} 
\affiliation{JEIP, UAR 3573, CNRS, Coll\`ege de France, PSL Research University, 11 Place Marcelin Berthelot, 75321 Paris Cedex 5, France}

\begin{abstract}
We investigate the nonequilibrium dynamics of the weak-coupling Hubbard-Holstein model after a sudden switch-on of the electron-phonon interaction within nonequilibrium dynamical mean-field theory (DMFT). Using the self-consistent Migdal approximation for the electron-phonon coupling together with second-order perturbation theory for the electron-electron interaction, we show that the relaxation dynamics exhibits a crossover between electron-dominated and phonon-dominated regimes, extending to finite Hubbard interaction the scenario previously identified in the Holstein model. To investigate the microscopic buildup of the thermal state, we analyze the dynamics within the Step-by-Step DMFT framework. In the plane of real time and DMFT iteration number, thermalization is marked by a sharp propagating front. 
This front appears in electronic observables already for weak quenches within the simulated time window, whereas the phononic sector exhibits a visible front only at sufficiently strong coupling. Thus, at weak coupling the local dispersionless phonons show a delayed onset of front formation, while near and beyond the crossover the front develops on comparable timescales in both the electronic and phononic sectors. 
Whenever both fronts are resolved, they propagate with the same velocity, showing that thermalization spreads coherently through the coupled electron-phonon system. 
\end{abstract} 
\maketitle 

%%%%%%%%%%%%%%%%%%%%%%%%%%%%%%%%%%%%%%%%%%%%%%%%%%%%%%%%%%%%%%%
\section{\label{sec:Introduction}Introduction}

The nonequilibrium dynamics of correlated lattice systems has been intensively investigated over the past decade, driven by rapid advances in experimental techniques that enable the manipulation of materials and control of their properties in the time domain~\cite{Basov2017}. Interaction-quench studies~\cite{Eckstein2008,Eckstein2009,Eckstein2010b,Aoki2014,Picano2021_AFM} have been strongly motivated by cold-atom experiments, where interactions and hopping amplitudes can be tuned via Feshbach resonances or by adjusting the depth of optical lattice potentials. In condensed-matter systems, ultrafast pump–probe experiments using wavelength-tunable femtosecond laser pulses allow for the driving of long-lived nonequilibrium states with novel functionalities that are inaccessible in equilibrium~\cite{Giannetti2016,delaTorre2021}. Such excitations can induce insulator-to-metal transitions in correlated Mott systems~\cite{Ogasawara2000,Perfetti2006} or even lead to ultrafast switching into metastable hidden phases absent from the equilibrium free-energy landscape~\cite{Stojchevska2014,Grandi2025_NOBD}. In many materials, electron–phonon coupling plays a crucial role in these processes, and pump–probe experiments often reveal clear signatures of coherent phonon dynamics~\cite{Huber2014,Maklar2021}. From a theoretical perspective, however, the interplay between electronic and lattice degrees of freedom out of equilibrium remains far from fully understood.

% %From Philipp and Yuta
% In Refs.~\cite{Vidmar2011,Golez2012,Golez2012b}, the
% nonequilibrium dynamics of one or two polarons in the Holstein model was investigated with a time-dependent exact diagonalization method.
% %From my paper SST
% Direct wave-
% function-based techniques like exact diagonalization and
% matrix product state algorithms~\cite{Jeckelmann1999,Ejima2009,Jansen2020} for electron-phonon
% coupled systems must cope with the large bosonic Hilbert
% space. 
% %From my paper SST
% Efficient procedures exist for the dilute limit of
% a few polarons~\cite{Vidmar2011,DeFilippis2012,Dorfner2015,Bonca1999}, but simulations at finite electron
% density~\cite{Sous2021,Jansen2021} remain restricted to short times. 
% %From Philipp and Yuta
% As for many-electron problems, previous
% works have investigated systems with classical phonons~\cite{Yonemitsu2009}
% or quantum phonons~\cite{Matsueda2012} in one dimension. Many-electron
% problems in two-dimensional systems have been studied with
% an exact diagonalization method for the Holstein-Hubbard
% model~\cite{DeFilippis2012} or with a weak-coupling perturbation theory for
% the Holstein model~\cite{Sentef2013,Kemper2013,Kemper2014}. 

Early theoretical studies of nonequilibrium electron–phonon systems have focused on few-particle problems, such as the dynamics of one or two polarons in the Holstein model, investigated using time-dependent exact diagonalization~\cite{Vidmar2011,Golez2012,Golez2012b}. More generally, wave-function-based approaches, including exact diagonalization and matrix product state methods~\cite{Jeckelmann1999,Ejima2009,Jansen2020}, are limited by the rapid growth of the bosonic Hilbert space. While efficient techniques exist in the dilute limit of a few polarons~\cite{Vidmar2011,DeFilippis2012,Dorfner2015,Bonca1999}, simulations at finite electron density remain restricted to relatively short times~\cite{Sous2021,Jansen2021}. Extensions to many-electron systems have been explored using classical phonons~\cite{Yonemitsu2009} or quantum phonons in one dimension~\cite{Matsueda2012}, as well as in two dimensions via exact diagonalization or weak-coupling perturbative approaches~\cite{DeFilippis2012,Sentef2013,Kemper2013,Kemper2014}.

In large spatial dimensions, nonequilibrium dynamical mean-field theory (DMFT)~\cite{Georges1996,Aoki2014}, which becomes exact in the limit of infinite dimensions, provides a controlled nonperturbative framework.
% provides a controlled and nonperturbative framework. DMFT maps the lattice problem onto an effective quantum impurity model subject to a self-consistency condition, thereby capturing local correlations exactly
DMFT maps the lattice problem onto an effective quantum impurity model subject to a self-consistency condition, thereby capturing local correlations exactly.
% %From Philipp and Yuta
% On the other hand, the dynamical
% mean field theory (DMFT)~\cite{Georges1996,Aoki2014}, which becomes exact in
% infinite spatial dimensions, has been used to study the interplay
% of electrons and phonons.
% From my paper SST
% For systems with large spatial dimension, non-equilibrium dynamical mean-field theory (DMFT) becomes the reference method. DMFT maps a lattice model with, e.g., local electron-phonon or electron-electron coupling to a single-site impurity
% model. In equilibrium, this model can be solved using quantum Monte Carlo (QMC) techniques~\cite{Assaad2007,Werner2007}, but
% non-equilibrium simulations usually rely on perturbative
% weak-coupling expansions~\cite{Murakami2015,Randi2017} or the strong coupling expansion around the atomic limit~\cite{Eckstein2010,Werner2013}. 
% Simulations in the Mott-insulating phase, which employ a strong-coupling impurity solver~\cite{Werner2013,Werner2015}, show that the feedback of the lattice dynamics on
% the electrons can lead to significant changes in the spectral
% function and to qualitatively different relaxation pathways. 
In equilibrium, this impurity problem can be solved using quantum Monte Carlo methods~\cite{Assaad2007,Werner2007}, whereas nonequilibrium studies typically rely on weak-coupling expansions~\cite{Murakami2015,Randi2017} or strong-coupling approaches around the atomic limit~\cite{Eckstein2010,Werner2013}. In the Mott-insulating regime, strong-coupling impurity solvers have revealed that the feedback of lattice dynamics on the electronic degrees of freedom can lead to significant modifications of spectral properties and to qualitatively distinct relaxation pathways~\cite{Werner2013,Werner2015}.

% In the weak-coupling Holstein model, nonequilibrium DMFT studies have  shown the existence of a crossover between two distinct relaxation regimes:
% In the weaker electron-phonon coupling
% regime, the phonon oscillations are damped faster than the
% thermalization time of the electrons, which contrasts with
% the stronger interaction regime where the relaxation of the
% phonons becomes slower than the electron relaxation~\cite{Murakami2015}.
% Despite these important findings, some important questions remains to be answered.
% It remains
% to be clarified how an electron-phonon system thermalizes microscopically in
% weakly or moderately correlated metallic systems~\cite{Murakami2015} and how
% the phonon dynamics affects the relaxation process beyond
% the conventional analysis based on the Boltzmann equation~\cite{Kabanov2008,Groeneveld1995}.  In addition, various interesting questions that have
% been addressed in purely electronic systems (such as the
% Hubbard model)~\cite{Eckstein2010b,Picano2021} remain to be answered for electron systems
% coupled to phonons
% in
% weakly or moderately correlated metallic systems.

In the weak-coupling Holstein model, where electrons couple to local (Einstein) phonons in the absence of Coulomb interactions, nonequilibrium DMFT studies have identified a crossover between two qualitatively distinct relaxation regimes~\cite{Murakami2015}. For weak electron–phonon coupling, phonon oscillations are damped faster than the electronic thermalization time, whereas for stronger coupling the phonon relaxation becomes the slowest process. Despite these advances, several fundamental questions remain open. In particular, the microscopic mechanisms governing thermalization in electron–phonon systems in weakly or moderately correlated metallic regimes are not yet fully understood, nor is the role of phonon dynamics beyond semiclassical or Boltzmann descriptions~\cite{Kabanov2008,Groeneveld1995}. Moreover, concepts of thermalization that have been extensively studied in purely electronic systems, such as the Hubbard model~\cite{Eckstein2010b,Picano2021}, remain largely unexplored in the presence of dynamical lattice degrees of freedom.

A central issue in this context is how an isolated quantum many-body system evolving under its own Hamiltonian approaches a thermal state and loses memory of its initial conditions~\cite{polkovnikov2011colloquium,Gogolin_2016,dalessio2016from}. Within DMFT, this problem acquires a particularly transparent interpretation: the lattice system effectively acts as its own bath, and thermalization emerges from the self-consistent feedback between the impurity and the dynamical bath generated by the surrounding lattice~\cite{Georges1996}. Recent work has shown that this mechanism can be elucidated by exchanging the order of the long-time limit and the DMFT self-consistency loop~\cite{Picano2025_thermalization}. In this Step-by-Step formulation, the impurity dynamics is first evolved to long times for a fixed bath, and the bath is subsequently updated iteratively. This approach reveals that, in the nonequilibrium Hubbard model, thermalization proceeds via a sharply defined front in the plane spanned by real time and DMFT iteration number, providing a microscopic picture of how a correlated system dynamically builds its own thermal environment.

An important open question is how this picture is modified when electronic and lattice degrees of freedom are dynamically coupled. In this work, we address this problem by studying the nonequilibrium dynamics of the weak-coupling Hubbard–Holstein model~\cite{Werner2007,Backes2023} after a sudden quench of the electron–phonon interaction. We consider both the  full nonequilibrium DMFT solution and  the Step-by-Step DMFT to investigate how thermalization emerges microscopically in a coupled electron–phonon system. This allows us to address two key questions: (i) whether the relaxation crossover identified in the Holstein model~\cite{Murakami2015} persists in the presence of finite Hubbard interactions, and (ii) how the self-consistent buildup of the thermal state manifests in electronic and phononic observables, and whether both sectors participate in a common propagating thermalization front.

The nonequilibrium DMFT impurity problem is solved using a conserving approximation that combines the self-consistent Migdal treatment of the electron–phonon interaction~\cite{Freericks1994,Bauer2011,Meyer2002,Capone2003,Koller2004b} with second-order perturbation theory for the electron–electron interaction~\cite{Tsuji2103b}. This framework captures the mutual feedback between electronic and lattice degrees of freedom and provides an ideal setting to investigate relaxation processes in isolated quantum many-body systems. In this way, the thermalizing system (electrons and phonons) is able to effectively act as its own thermal bath~\cite{Nandkishore2015}, enabling it to thermalize its subsystems and to lose memory of its initial conditions.

We find that the crossover between electron- and phonon-dominated relaxation remains robust in the weak-coupling regime of the Hubbard–Holstein model. The Step-by-Step analysis further reveals a pronounced asymmetry in the onset of thermalization: within the accessible simulation times, electronic observables exhibit a well-defined thermalization front already for weak quenches, whereas the phononic sector develops a clear front only at sufficiently strong electron–phonon coupling. Whenever both fronts are present, they propagate with the same velocity, demonstrating that thermalization spreads coherently throughout the coupled electron–phonon system. These findings provide a microscopic picture of how thermal equilibrium emerges self-consistently in interacting electron–phonon systems within DMFT.

The paper is organized as follows. In Sec.~\ref{sec:model}, we introduce the model and the nonequilibrium DMFT formalism~\cite{Murakami2015}. In Sec.~\ref{sec:ResultsPt1}, we analyze the relaxation dynamics and the crossover between electron- and phonon-dominated regimes~\cite{Murakami2015}. In Sec.~\ref{sec:ResultsPt2}, we investigate the emergence of thermalization within the Step-by-Step DMFT framework~\cite{Picano2025_thermalization}. In Sec.~\ref{sec:ResultsPt3}, we study the dependence of the thermalization front velocity on the model parameters. Finally, Sec.~\ref{sec:Conclusion} contains our conclusions.

%%%%%%%%%%%%%%%%%%%%%%%%%%%%%%%%%%%%%%%%%%%%%%%%%%%%%%%%%%%%%%%%%%%%%%%%%%%%%%%%%%%%%%%%%%%%%%%%%%%%%%

\section{Model }
\label{sec:model}

\subsection{Nonequilibrium DMFT for the Hubbard-Holstein model}
The Hamiltonian for the Hubbard-Holstein model is
\begin{align}
\label{eqn:Hamilt}
\mathcal{H}(t) &= -v \sum_{\langle i,j \rangle, \sigma} (c_{i,\sigma}^\dagger c_{j,\sigma} 
+ \text{h.c.}) - \mu \sum_i n_i \\
\nonumber
&+ \omega_0 \sum_i a_i^\dagger a_i 
+ g(t) \sum_i (a_i^\dagger + a_i)(n_{i,\uparrow} + n_{i,\downarrow} - \alpha) \\ 
\nonumber
&+ U\sum_{i} \left(n_{i\uparrow}-\frac{1}{2}\right)\left(n_{i\downarrow}-\frac{1}{2}\right),
\end{align}
where: $c_{i,\sigma}^\dagger$ is the creation operator for an electron with spin $\sigma$ on site $i$; $v$ is the hopping parameter restricted to next-neighboring sites; $a^\dagger$ is the creation operator for a phonon with frequency $\omega_0$; $g(t)$ is the  electron-phonon coupling strength, taken here to be time-dependent; and $\alpha$ is a constant that can be chosen arbitrarily. In the following, we set $\alpha = \langle n_\uparrow + n_\downarrow \rangle$,  so that the Hartree term in the self-energy vanishes.
We assume the absence of long-range order and focus on the half-filled case. Since we are at half-filling, $\alpha= \langle n_\uparrow + n_\downarrow \rangle=1$. We set $\mu=0$ in order for the system to be particle-hole symmetric. $U$ denotes  the Coulomb interaction energy. 
% In the following, we consider a Bethe lattice, with hopping $v=1$ setting the unit of energy and semi-elliptic density of states $D(\epsilon)=\sqrt{4-\epsilon^2}/(2\pi)$.
% We also note that in the anti-adiabatic limit ($\omega_0 \to \infty$ with $\lambda = 2g^2/\omega_0$ fixed) the Holstein model becomes the attractive Hubbard model with a non-retarded interaction $-\lambda$.
Finally, we introduce the position ($X$) and momentum ($P$) operators for the phonons,
\begin{align}
X_i &= \frac{1}{\sqrt{2}} (a_i^\dagger + a_i), \\
P_i &= \frac{i}{\sqrt{2}} (a_i^\dagger - a_i).
\end{align}
If we substitute these expressions for $X$ and $P$ (we are in adimensional units) into the Hamiltonian \eqref{eqn:Hamilt}, we get:
\begin{align}
    &\mathcal{H}_{\text{free-ph},i} = \omega_0 a_i^\dagger a_i = \frac{\omega_0}{2} (X_i^2+P_i^2) - \frac{\omega_0}{2}\nonumber \\
    &\mathcal{H}_{\text{el-ph},i} =g(t) (a_i^\dagger + a_i)(n_{i,\uparrow} + n_{i,\downarrow} - \alpha) = \sqrt{2} g(t) X_i (n_i - \alpha)
\end{align}
where in the first equation we used $[X,P]=i$.
 We drive the system out of equilibrium by changing the electron-phonon coupling constant $g(t)$ from $g_i=0$ to a finite value $g_f$ at $t=0^+$.
 % or by coupling the system with a photoexcitation bath at negative temperature.

To investigate the out-of-equilibrium dynamics of the Hubbard-Holstein model after the quench, we solve nonequilibrium dynamical mean-field theory (DMFT)
~\cite{Aoki2014,Murakami2015} equations on the L-shaped Kadanoff-Baym contour $\mathcal{C}$~\cite{KadanoffBaym1962}.
% , which runs from $t = 0$ up to the maximum simulation time $t_{\text{max}}$ along the real-time axis, back to $t = 0$, and then proceeds to $-i\beta$ along the imaginary-time axis, where $\beta = 1/T$ is the inverse temperature of the initial equilibrium state. 
We define the lattice electronic Green's function $G_{i,j,\sigma}(t, t')$ and  local phonon Green's function $D_i(t,t')$ on this contour as
\begin{align}
G_{i, j, \sigma}(t,t') &= -i\langle T_\mathcal{C} c_{i,\sigma}(t) c_{j,\sigma}^\dagger(t') \rangle, \\
D_i(t,t') &= -2i \langle T_\mathcal{C} X_i(t) X_i(t') \rangle,
\end{align}
where $T_\mathcal{C}$ is the contour-ordering operator.
The DMFT formalism assumes a spatially local self-energy ($\Sigma_{i,j} (t,t')\equiv \delta_{i,j}\Sigma_{ii}(t,t')$) and maps the lattice problem onto a quantum impurity model in a self-consistent manner.
% DMFT becomes exact in the limit of infinite coordination number ($z \to \infty$) where, in order to evaluate the correct functional $\Sigma_{ii} [G]$, it is sufficient to solve a local model with action:
The effective impurity action for the Hubbard-Holstein model is:
\begin{align}
\label{eqn:Simp}
S_{\text{imp}} &= -i \int_{\mathcal C} dt\, H_{\mathrm{loc}}(t) \nonumber \\
     &- i \sum_\sigma \int_{\mathcal C} dt \int_{\mathcal C} dt'\,
     c_\sigma^\dagger(t)\,
     \Delta_{ \sigma}(t,t')\,
     c_\sigma(t')
\end{align}
where 
% the auxiliary field $\Delta(t,t')$ is chosen such that 
% \begin{align}
% G_{ii,\sigma}(t,t') 
% = -i \langle \mathcal T_{\mathcal C} \, c_\sigma(t) c_\sigma^\dagger(t') \rangle_{S_i}
% \end{align}
% and $\Sigma$ is implicitly defined via the Dyson equation
% \begin{align}
% G_{ii,\sigma}^{-1}(t,t') 
% = (i\partial_t + \mu)\,\delta_{\mathcal C}(t,t')
% - \Sigma_{ii,\sigma}(t,t')
% - \Delta_{i,\sigma}(t,t')
% \end{align}
\begin{align}
H_{\mathrm{loc}}(t)
&= -\mu (n_\uparrow + n_\downarrow) + 
U (n_\uparrow-\frac{1}{2}) (n_\downarrow - \frac{1}{2}) \nonumber \\
&+ \omega_0 a^\dagger a 
+ g(t) (a^\dagger + a)\,(n_\uparrow + n_\downarrow - \alpha)
\end{align}
The hybridization function $\Delta$ is determined self-consistently in such a way that the
electron Green’s function for the impurity ($G_{\text{imp}}$) becomes
identical to the local electron Green’s function of the lattice
$G_{\text{loc}} \equiv G_{i,i,\sigma}$, where the self-energy of the lattice system is identified with that of the effective impurity problem (self-consistency condition). 
A closed relation $\Delta[G]$ can be obtained, for example, 
in the Bethe lattice with infinite coordination number
($z \to \infty$), where the self-consistency condition simplifies to
\begin{align}
    \label{eqn:Delta}
    \Delta_{\sigma} (t,t') = v_*^2 G_{\text{loc},\sigma} (t,t')
\end{align}
with $v=v_*/\sqrt{z}$. In this case, that we consider in this paper, the density of states is semi-elliptic, $\frac{1}{2 \pi v_*^2} \sqrt{4 v_*^2 - \epsilon^2}$, and we set $v_* = 1$ in the following. Especially, the band-width is $4$.

Since the electrons interact with each other through the phonons and vice versa, to obtain the interacting Green’s functions, we introduce the self-energies $\Sigma_\sigma (t,t')$ and $\Pi(t,t')$ for the electrons and phonons, respectively. These functions satisfy the Dyson equations,
\begin{align}
G_{\text{imp},\sigma}(t,t') &= \mathcal{G}_{0,\sigma}(t,t') + \left[ \mathcal{G}_{0,\sigma} \ast  \Sigma_\sigma \ast  G_{\text{imp},\sigma} \right](t,t'), \label{eqn:G_imp_Dyson} \\
D(t,t') &= D_0(t,t') + \left[ D_0 \ast \Pi \ast D \right](t,t') \label{eqn:D_Dyson}
\end{align}
where $\ast$ denotes convolution on the contour $\mathcal{C}$, and
$\mathcal{G}_{0,\sigma}$ is the Weiss Green’s function for the impurity problem, which is related to the hybridization function $\Delta_\sigma (t,t')$ by:
\begin{align}
    \mathcal{G}_{0,\sigma}^{-1} (t,t') = (i \partial_t + \mu)\delta_{\mathcal{C}}(t,t') - \Delta_\sigma (t,t')\,,
\end{align}
Because the phonon propagator is local and site independent in the homogeneous phase, we drop the site index and denote it by \(D(t,t')\).
The electronic polarizability is defined as 
\begin{align}
    \Pi(t,t') &= - i 2 g(t) g(t') G_{\text{imp}}(t, t') G_{\text{imp}}(t', t)
\label{eqn:Polariz1}
\end{align}
while the non-interacting ($g=0$) phonon Green's function can be expressed as
\begin{align}
D_0(t,t') &= -i \left[ \theta_\mathcal{C}(t,t') + f_B(\omega_0) \right] \exp\left( -i\omega_0 \int_{\mathcal{C},t'}^t dt_1 \right) \nonumber \\
&\quad -i \left[ \theta_\mathcal{C}(t',t) + f_B(\omega_0) \right] \exp\left( i\omega_0 \int_{\mathcal{C},t}^{t'} dt_1 \right),
\end{align}
where $\theta_\mathcal{C}$ is the Heaviside function on the contour and $f_B(\omega_0) = (e^{\beta\omega_0} - 1)^{-1}$ is the Bose distribution function at inverse temperature $\beta = 1/T$.
In the paramagnetic phase, the impurity Green's function is spin independent, \(G_{\mathrm{imp},\uparrow}=G_{\mathrm{imp},\downarrow}\equiv G_{\mathrm{imp}}\), and the factor \(2\) in the polarizability accounts for the spin sum.

\subsection{Impurity Solvers}

The most demanding step in the DMFT self-consistency loop is the solution of the effective impurity problem, Eq.~\eqref{eqn:Simp}. A numerically exact solution is in principle to be obtained with quantum Monte Carlo (QMC) methods, as in the Hubbard model~\cite{Eckstein2009,Eckstein2010b}. However, such methods are formulated on the real axis, which makes it difficult to access the long times required to simulate phonon dynamics. Therefore, we employ here two  approximate diagrammatic impurity solvers (weak-coupling approximations): the self-consistent second-order perturbation theory for the electronic problem~\cite{Tsuji2013}, and the self-consistent Migdal approximation~\cite{Randi2017} for the phononic problem.

\subsubsection{Electronic problem: Second-order perturbation theory}
In this work, we focus on the weakly correlated electronic regime.
We employ the self-consistent second-order perturbation
theory (SPT) as an impurity solver~\cite{Tsuji2013} which, being a
conserving approximation, is particularly suited to studying
thermalization in isolated quantum systems and the energy
transfer between the impurity and the bath at weak coupling~\cite{Aoki2014}. 
Since SPT 
is a “conserving approximation”~\cite{Baym1961}, it automatically guarantees the conservation of global quantities such as the total
energy and the particle number. 
The perturbation theory defines the self-energy as a functional of the fully interacting $G$, $\Sigma = \Sigma [G]$, which is a sufficient condition to preserve the conservation laws:
    \begin{align}
	   \label{eqn:SPT}
       \Sigma_{\sigma}^{\text{e-e}}(t,t') = U(t) U(t')\, G_{\text{imp},\sigma}(t,t')\, G_{\text{imp},\bar\sigma}(t',t)\, G_{\text{imp},\bar\sigma}(t,t')
    \end{align}
% In this 
% manuscript, we  consider $U/(4v_\ast) = 0, 0.25, 0.5$ in the weak-coupling regime ($v_\ast = 1$).
The dynamics in the strongly correlated regime would
require a different approach, such as for the example the
noncrossing approximation~\cite{Eckstein2010}. 
% In SPT, the self-energy diagrams are expanded with respect to the fully interacting Green’s function $G_{\text{imp}}$:
In the following, we will drop the spin index since we will  always be in the paramagnetic case.

% It
% is important that the conservation law is satisfied in a simulation of the time evolution to obtain physically meaningful
% results. 

\subsubsection{Phonon problem: Self-consistent Migdal approximation}
 In the Migdal approximation, one treats within a Gaussian approximation the quantum fluctuations of the phonons around the average order parameter, represented by the displacement $\langle X \rangle$ of the atoms~\cite{Randi2017}. 
The self-consistent Migdal approximation is the lowest order approximation for the electronic self-energy that allows to treat the renormalization of the phonons induced by the electron-phonon coupling. The vibrational mode evolves as a consequence of the interaction with the electrons and, in turn, influences the electronic dynamics; the electrons couple back to the phonons in the form of a phonon self-energy (polarization bubble). 
% The expectation values that will be shown in the following denote expectation values of quantum operators.
The self-energy in the self-consistent Migdal approximation reads:
\begin{align}
\label{eqn:SigmaMigdal}
% &\Sigma(t,t') = \nonumber \\
% &- \delta_c(t,t') g(t) \int dt_1 \left[ \alpha + 2i G_{\text{imp}}(t_1, t_1 + 0_{\mathcal{C}}^+) \right] D_0(t_1, t) g(t_1) \nonumber \\
% &+ i D(t, t') G_{\text{imp}}(t, t') g(t) g(t'),
\Sigma^{\text{el-ph}}(t,t') &=  i D(t, t') G_{\text{imp}}(t, t') g(t) g(t')
\end{align}
% This approximation has been used to study the Holstein model in equilibrium, and its accuracy has been discussed in a number of papers.\textsuperscript{48,50–54}.
As long as $g$ is not close to the critical value $g_c$ for the transition to the bipolaronic insulating phase  and $g$ is small compared to the electron bandwidth (both conditions that will be fulfilled in this manuscript), it provides a qualitatively good description~\cite{Murakami2015,Bauer2011,Koller2004}. Since the self-energies of the electrons and phonons involve dressed propagators, we can take account of the interplay between the electrons and phonons in the dynamics. With the choice of $\alpha = \langle n_\uparrow + n_\downarrow \rangle$ in Eq.~\eqref{eqn:Hamilt}, the Hartree term (not shown in Eq.~\eqref{eqn:SigmaMigdal}) vanishes, and we can define a Luttinger-Ward functional $\Phi[G, D]$ in this approximation. Hence the Migdal approximation is a conserving one.
Since both SPT and self-consistent Migdal correspond to skeleton truncations of the Luttinger–Ward functional, the combined approximation within DMFT is $\Phi$-derivable and therefore conserving~\cite{Baym1962}.

% \subsection{Hartree-Fock approximation}

% As we mentioned in the introduction, the Hartree-Fock (HF) approximation is also sometimes called the (unrenormalized) Migdal approximation.\textsuperscript{49} In this approximation, the electron self-energy 
% is given by
% \begin{align}
% &\Sigma(t,t') = \nonumber \\
% &- \delta_c(t,t') g(t) \int dt_1 \left[ \alpha + 2i G_{\text{imp}}(t_1, t_1 + 0_{\mathcal{C}}^+) \right] D_0(t_1, t) g(t_1) \nonumber \\
% &+ i D_0(t, t') G_{\text{imp}}(t, t') g(t) g(t'),
% \end{align}
% The Feynman diagrams for the self-energy have the same structure as in the self-consistent Migdal approximation, but the dressed phonon propagator is replaced with the bare equilibrium one $D_0$. Thus, in the HF approximation, we ignore the phonon self-energy, which means there is no feedback from the electrons to the phonons (Fig.~1(b)). Hence we cannot extract the dynamics for the phonons from this scheme. Also, the HF approximation cannot be derived from a Luttinger-Ward functional, and is thus not conserving.

% The HF approximation has been used to describe equilibrium states\textsuperscript{49} and nonequilibrium dynamics\textsuperscript{36} for the Holstein model. In addition, the HF scheme for small $g$ has been used in some DMFT studies to describe the effect of a bosonic heat bath on the electrons.\textsuperscript{57,58} The results in Section IIIC will confirm that the phonon effectively acts as heat bath within the HF approximation.

\subsection{Full DMFT scheme}
% This section discusses the solution of the %Anderson-Holstein impurity model
% effective impurity problem
% with the self-consistent Migdal approximation \cite{Randi2017}.
% In the Migdal approximation, one treats within a Gaussian approximation the quantum fluctuations of the phonons around the average order parameter, represented by the displacement $\langle X \rangle$ of the atoms. 
% The self-consistent Migdal approximation is the lowest order approximation for the electronic self-energy that allows to treat the renormalization of the phonons induced by the electron-phonon coupling. The vibrational mode evolves as a consequence of the interaction with the electrons and, in turn, influences the electronic dynamics; the electrons couple back to the phonons in the form of a phonon self-energy (polarization bubble). The expectation values that will be shown in the following denote expectation values of quantum operators.
Here we summarize the self-consistent DMFT loop that we solve for each time on the contour $\mathcal{C}$.

\begin{enumerate}
    \item Start from a guess for $\Sigma(t,t')$ and evaluate the fully-interacting impurity Green's function $G_{\text{imp},\sigma}$ (that in the following will be simply $G$) by solving the corresponding Dyson equation:
    \begin{align}
\label{eqn:Dyson_G}
&(i \partial_t + \mu)G(t,t') \nonumber \\
&- [\Delta(t,t')+\Sigma(t,t')] * G(t,t') =\mathcal{\delta}_{\mathcal{C}}(t,t')
\end{align} 
%     \begin{align}
% \label{Dyson_G}
% &[i \partial_t + \mu 
% -h_\text{loc}(t)
% ]G(t,t') \nonumber \\
% &- [\Delta(t,t')+\Sigma(t,t')] * G(t,t') =\mathcal{\delta}_{\mathcal{C}}(t,t')
% \end{align} 
in the Keldysh formulation  (following the notation for two-time Green's functions in Ref.~\cite{Aoki2014}). We are considering here the spin-symmetric phase and omitting spin indices in  $\Sigma$, $G$ and $\Delta$.
    \item Use the self-consistency Eq.~\eqref{eqn:Delta} to fix the hybridization function of the  impurity model:
\begin{align}
    \Delta (t,t') = v_*^2 G (t,t')
\end{align}
    \item Solve the impurity model to get $\Sigma^{\text{el-ph}} [G, D]$. We implement the self-consistent Migdal approximation for the electron-phonon interaction:
    \begin{enumerate}
        \item Calculate the phonon self-energy $\Pi$ (polarization operator) with Eq.~\eqref{eqn:Polariz1} in order to include the back-action of the electrons on the phonons:
        \begin{align}
\Pi(t,t') &= - i 2 g(t) g(t') G(t, t') G(t', t)
\end{align}
        \item Solve the Dyson equation for the dressed phonon propagator $D$, Eq.~\eqref{eqn:D_Dyson}:
        \begin{align}
        [1- D_0(t,t')* \Pi(t,t')]* D(t,t') = D_0(t,t')
        \end{align}
        \item  Calculate the second-order electronic self-energy in the self-consistent Migdal approximation, Eq.~\eqref{eqn:SigmaMigdal}:
    \begin{align}
        \Sigma^{\text{el-ph}}(t,t') &=  i D(t, t') G(t, t') g(t) g(t'),
    \end{align}
\end{enumerate}
    \item Solve the impurity model for the electron-electron interaction to get $\Sigma^{\text{e-e}} [G]$. With SPT as an impurity solver, we apply Eq.~\eqref{eqn:SPT}: 
    \begin{align}
\Sigma^{\text{e-e}}(t,t') = U(t) U(t')\, G(t,t')\, G(t',t)\, G(t,t')
    \end{align}
    \item Set $\Sigma(t,t') = \Sigma^{\text{e-e}}(t,t') + \Sigma^{\text{el-ph}}(t,t')$  and iterate steps 1.-- 5. until convergence. 
    \end{enumerate}

In the self-consistent Migdal approximation the vibrational mode evolves as a consequence of the interaction with the electrons. The expectation value of $X$ is determined by the exact equation of motion that comes form
\begin{align}
    \dot X &= \frac{\partial \mathcal{H}}{\partial P } = \omega_0 P = \omega_0 \frac{i}{\sqrt{2}} (a^\dagger - a) \nonumber \\
    \dot P &= -\frac{\partial \mathcal{H}}{\partial X } = -\omega_0 X - \sqrt{2} g(t) (n-1) 
\end{align}
and, if we take the expectation value, it reads:
\begin{align}
    \label{eqn:Xdotodot}
    \langle \ddot X(t) \rangle  
    &= \omega_0  \langle \dot P(t) \rangle \nonumber \\ 
    &=   -\omega_0^2  \langle X(t) \rangle  - \sqrt{2} \omega_0 g(t) \big (  n(t) -1 \big )
\end{align}
where $n(t) = -i G^<(t,t)$. 
We see explicitly from Eq.~\eqref{eqn:Xdotodot} that dynamics of the phonon is influenced by the electronic dynamics through the electron density. The latter, in turn, depends on the average position of the phonon displacement $\langle X(t) \rangle$  through the terms $\Delta$ and $\Sigma$ in Eq.~\eqref{eqn:Dyson_G}. 

\subsubsection{Observables}
The total energy of the system is given by the combination of different energy contributions.

\textit{Kinetic energy}: The kinetic energy of the electrons is calculated as:
\begin{align}
E_{\text{kin}}(t) &= -i \sum_{\sigma} \left [ \Delta_{\sigma} \ast G_{\text{imp},\sigma} \right] ^<(t,t),
\label{eqn:Ekin}
\end{align}

\textit{Phonon density}:
The density of phonons $\langle a^\dagger(t)a(t)\rangle$, which is proportional to the free-phonon energy \(E_{\mathrm{ph}}(t)=\omega_0\langle a^\dagger(t)a(t)\rangle\),
can be expressed in terms of the $X$ and $P$ as
\begin{align}
\label{eqn:phononDensity}
\langle a^\dagger(t) a(t) \rangle
&= \frac12
   \left[\langle X(t)X(t)\rangle +
  \,\langle P(t)P(t)\rangle \right ]
   - \frac12
\end{align}
where $\langle X(t) X(t) \rangle$ is obtained from $D(t,t)$, while $\langle P(t) P(t) \rangle$ is calculated from a second derivative of $D(t, t')$.
In particular, the variance of the displacement $X$ is extracted from: $\text{Var}[X] \equiv \langle \Delta X^2 \rangle = \frac{i}{2} D^<(t,t)$, while $\text{Var}[P] \equiv \langle \Delta P^2 \rangle = \frac{i}{2} D_{d1,d2}^<(t,t)$. The second derivative of $D$ is defined as:
\begin{align}
    D_{d1,d2}(t,t')\equiv \frac{\partial_t \partial_{t'} D(t,t')}{\omega_0^2} = \frac{2}{\omega_0} \delta_{\mathcal{C}}(t,t') - 2 i \langle T_{\mathcal{C}} P(t)P(t') \rangle
\end{align}
(See \cite{Murakami2015} for additional reference.)

\textit{Electron-phonon correlation}: The energy term due to electron-phonon interaction is:
\begin{equation}
\label{eqn:elphoncorr}
\sqrt{2} g(t) \langle X(t) c^\dagger_{\sigma}(t) c_{\sigma}(t) \rangle = -i \left[\Sigma^{\text{el-ph}} * G_{\text{imp},\sigma} \right]^<(t, t),
\end{equation}

\textit{Electron-electron correlation}: The energy term due to Hubbard electron-electron interaction is:
\begin{align}
E_U(t)
&= U(t)\Big\langle
n_{\bar\sigma}(t)
\left(n_{\sigma}(t)-\frac12\right)
\Big\rangle \nonumber \\ 
&=
-i\,[\Sigma_\sigma * G_{\text{imp},\sigma}]^{<}(t,t)
-
\frac{U(t)}{2}
\left\langle
n_{\sigma}(t)-\frac12
\right\rangle
\end{align}
at half-filling the second term vanishes and
\begin{align}
E_U(t)
=
-i\,[\Sigma_\sigma * G_{\text{imp},\sigma}]^{<}(t,t)
\end{align}

\textit{Total energy}:
The total energy is given by:
\begin{align}
E_{\mathrm{tot}}(t)
&=
E_{\mathrm{kin}}(t)
+
E_U(t)
-
\mu \,\langle n(t)\rangle
+
\omega_0 \langle a^\dagger(t)a(t)\rangle \nonumber\\
&\quad+
\sqrt{2}\,g(t)\left[
\langle X(t)n(t)\rangle-\alpha \langle X(t)\rangle
\right].
\end{align}
with $n(t)=n_{\uparrow}(t)+n_{\downarrow}(t)$, where at half-filling we remind that $\mu=0$ and $\alpha=1$.

% \textit{Symmetry-broken solution --} In passing, we mention that in the symmetry-broken phase, the coordinate $X$ acquires a nonzero expectation value. In the case of a two-sublattice symmetry broken phase, we have two inequivalent impurity models 
% \begin{align}
% S_a &= -i \sum_{\sigma} \int_{\mathcal{C}} dt \left[ \sqrt{2} g X_a(t) \left( c^\dagger_{\sigma,a} c_{\sigma,a} - \frac{1}{2} \right) + \frac{\omega_0}{2} \left( X_a^2 + P_a^2 \right) \right]  \nonumber \\
% &- i \sum_{\sigma} \int_{\mathcal{C}} dt_1 dt_2 \, c^\dagger_{\sigma,a}(t_1) \Delta_a(t_1, t_2) c_{\sigma,a}(t_2) \,,
% \end{align}
% which represent sites on the $a$ and $b$ sublattice, i.e., all quantities, $G$, $\Delta$, $h_{\text{loc}}$, $\langle X(t) \rangle$, $\Sigma$, $P$, will additionally depend on the sublattice $a$ and $b$.  
% For the particle - hole symmetric case, we have $\langle X(t) \rangle_a = -\langle X(t) \rangle_b$. We use a bipartite lattice with a semielliptic density of states, in which the DMFT self-consistency is given by, 
% \begin{align}
% \Delta_a(t, t') &= v(t) G_b(t, t') v(t') \nonumber \\
% \Delta_b(t, t') &= v(t) G_a(t, t') v(t').
% \end{align}
% where $v(t)$ is the time-dependent profile of the hopping amplitude. This closes the DMFT equations. 

\subsection{Step-by-step DMFT}

% To understand the emergence of thermalization within DMFT, we will take a different perspective and look at the problem from the point of view of the impurity as its quantum bath is self-consistently updated~\cite{Picano2025_thermalization}. 
% Instead of converging the DMFT self-consistency at each time step, we proceed by first evolving the dynamics of the impurity $G(t,t')$ at fixed bath $\Delta(t,t')$ up to long times and then imposing the self-consistency after each iteration $n$. This amounts in practice in exchanging the order of limits between $t\rightarrow\infty$ (long-time limit) and $n\rightarrow\infty$ (DMFT self-consistency). We called this approach Step-by-Step DMFT~\cite{Picano2025_thermalization}.  

In the Step-by-Step DMFT, we start from the converged equilibrium solution of the DMFT system of equations at the initial temperature $T_i=1/\beta_i$ as in the full-DMFT case.
In the first global-DMFT iteration, $n=0$, we propagate the equilibrium solution up to $t_{\text{max}}$ without applying any excitation ($g_{n=0}=0,\,\,\forall t$). 
%For the very first DMFT iteration, $n=0$,  we do not apply any quench in the electron-phonon coupling $g$: $g=0  \, \,\,\,\forall t$. In this way the system is not excited and stays at the initial equilibrium temperature for all the times.
% Having reached self-consistency at equilibrium, we propagate the solution by means of the usual Dyson's equations~\cite{Aoki2014} for all $(t,t')$.
% Here we reach DMFT-convergence for each $(t,t')$, as we did in the normal approach, the only difference being the fact that the excitation is always off, so the system is in equilibrium during the whole dynamics.
% Actually, we are imposing that $\Delta_{n=0}(t,t')$, $G_{n=0}(t,t')$, $\Sigma_{n=0}(t,t')$, $\Pi_{n=0}(t,t')$ and $D_{n=0}(t,t')$ are time translational invariant (TTI) functions in all their components.
For DMFT-iterations $n=1,2,\dots, n_{\text{max}}$, we switch on the excitation by making the quench on $g$ at $t=0^+$, $g_n: 0 \to g_f$, and propagate the Green's functions in time up to $t_{\rm max}$ without imposing the DMFT self-consistency. 
% Actually, for each time $t$, we solve the equation Eq.~\eqref{eqn:Xdotodot} for the update of $\langle X_n(t) \rangle$,
% \begin{align}
%     \langle \ddot X_n(t) \rangle   
%     =   -\omega_0^2  \langle X_n(t) \rangle  - \sqrt{2} \omega_0 g(t) \big ( \langle n_n(t) \rangle -1 \big ) \, , 
% \end{align}
% and, given the updated $\langle X_n (t)\rangle$, 
For each $(t,t')$ on the contour $\mathcal{C}$, we solve \textit{only once} the DMFT equations (without imposing any self-consistency):
\begin{equation}
    \begin{cases}
&(i \partial_t + \mu) G_n(t,t') \nonumber \\ 
&-[\Delta_n(t,t')+\Sigma_n(t,t')] \ast G_n(t,t') =\mathcal{\delta}_{\mathcal{C}}(t,t')
\nonumber \\    
% &h_{\text{loc},n}(t)=\sqrt{2}g(t) \langle \hat X_n(t) \rangle + U(t) (n(t)-\frac{1}{2}) \nonumber \\ 
&\Delta_n(t,t')= v_\ast(t) G_{n-1}(t,t') v_\ast(t')  \nonumber \\
&\Pi_n(t,t')= - 2i g_n^2(t) G_n(t,t') G_n(t',t) \nonumber \\
&[1- D_0(t,t') \ast \Pi_n(t,t')] \ast D_n(t,t') = D_0(t,t') \nonumber \\
&\Sigma_n^{\text{el-ph}}(t,t')= i g_n^2(t) G_n(t,t') D_n(t,t') \nonumber \\
 &\Sigma_{n}^{\text{e-e}}(t,t') = U(t) U(t')\, G_{n}(t,t')\, G_{n}(t',t)\, G_{n}(t,t')
 \nonumber \\
 &\Sigma_{n}(t,t') = \Sigma_{n}^{\text{e-e}}(t,t') + \Sigma_{n}^{\text{el-ph}}(t,t')
    \end{cases}
\end{equation}
 We impose that \textit{global} convergence is reached for $n=n_{\text{max}}$ when
 \begin{align}
 	\sum_{t,t'} \abs{G_{n_{\text{max}}}(t,t')-G_{n_{\text{max}}-1}(t,t')}< 10^{-6}
 \end{align}
where $10^{-6}$ is the error threshold that we arbitrarily fix and the sum runs over all the times.

 We observe that the hybridization function $\Delta_n(t,t')$ depends on the $G_{n-1}(t,t')$ at the previous DMFT-iteration:
\begin{align}
\label{eqn:Deltan}
    \Delta_n(t,t')= v_\ast(t) G_{n-1}(t,t') v_\ast(t')  
\end{align}
 
 The crucial property of the bath $\Delta_n (t,t')$ is that, at finite $n$, it remains explicitly non-time-translation-invariant and retains memory of the initial condition.
 %This is due to the fact that $G_{n-1}$ keeps the memory of the initial temperature $T_i$.

%%%%%%%%%%%%%%%%%%%%%%%%%%%%%%%%%%%%%%%%%%%%%%%%%%%%%%%%%%%%%%%
\section{\label{sec:Results} Results and discussion} 
\subsection{\label{sec:ResultsPt1}Thermalization crossover in the Hubbard-Holstein model}

\begin{figure*}
    \centering
    \includegraphics[width=\linewidth]{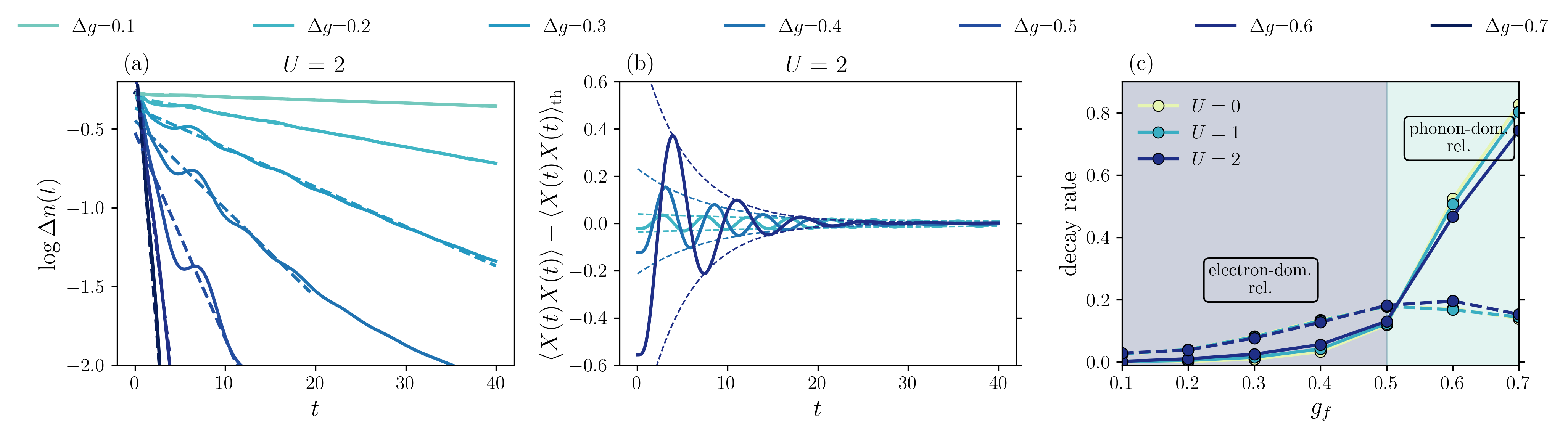}
    \caption{(a) Temporal evolution of $\Delta n (t)$ on logarithmic scale for various values of $g_f$ after the quench $g: 0 \to g_f$ at $t=0^+$ at $U /v_\ast=2$. Dashed lines are exponential fits. (b): temporal evolution of $\langle XX \rangle -\langle XX \rangle_{\text{th}} $ at  $U /v_\ast=2$. The dashed lines show exponential fits to the envelopes of the oscillating curves.
    (c) Electron (continuous lines) and phonon (dashed lines) energy scales (inverse relaxation times) as a function of $g_f$, extracted from panels (a) and (b), respectively, for $U / v_\ast = 0, 1, 2$. For $g_f < 0.5$, we observe electron-dominated relaxation dynamics; for $g_f > 0.5$ phonon-dominated relaxation.}
    \label{fig:Fig1}
\end{figure*}

In what follows, we consider the half-filled Hubbard-Holstein model with \(\omega_0=0.7\) (\(<W=4\), with \(W\) the bandwidth), in the weak-coupling  regime far from both the bipolaronic and Mott transitions. 
% Specifically, the electron-phonon coupling remains well below the critical value \(g_c\) for the bipolaronic transition, for which CT-QMC calculations yield \(0.8 \lesssim g_c \lesssim 0.85\) for \(10<\beta<40\), while the Hubbard interaction is restricted to \(U/W=0,\,0.25,\,0.5\), i.e. well below the Mott critical interaction.
We focus on $U/v^\ast = 0, 1, 2$
(which corresponds to weak interaction on the scale of the bandwidth, $U/W = 0, 0.25, 0.5$).
The system is initially prepared at \(g=0\) in thermal equilibrium at inverse temperature \(\beta=100\).
We then study the nonequilibrium dynamics induced by a sudden quench of the electron-phonon coupling, \(g:0\to g_f\), at \(t=0^+\). The corresponding \(U=0\) case, i.e. the Holstein model, was analyzed in Ref.~\cite{Murakami2015}; here we focus on how this relaxation scenario is modified by finite Hubbard interaction.

Figure~\ref{fig:Fig1} summarizes the relaxation dynamics. Panel (a) shows \(\Delta n(t)\) for \(U/v_\ast=2\) and several values of the final coupling \(g_f\).
Here \(\Delta n(t)\) quantifies 
the jump in the momentum distribution \(n(\epsilon_{\mathbf{k}},t)=-iG_{\mathbf{k}}^{<}(t,t)\)  at the Fermi level, \(\epsilon=0\).
Because the initial state is at low but finite temperature and may already include finite Hubbard interaction, one has $\Delta n(0)\lesssim 1$, as also visible from the equilibrium curve at $\beta_i$ in Fig.~\ref{fig:Fig2} (a)).
The quench injects energy into the system, so that the  thermal state  associated with the conserved total energy after the quench corresponds to a higher effective temperature. 
The resulting reduction of \(\Delta n(t)\) therefore provides a direct measure of electronic thermalization. For all quenches, \(\Delta n(t)\) shows an approximately exponential decay over a sizable time window, with superimposed oscillations, as indicated by the dashed fits. The decay becomes faster as \(g_f\) increases: for weak quenches the jump remains finite up to the longest accessible times, while for larger \(g_f\) it is rapidly suppressed. Hence, similarly to the Hubbard model~\cite{Eckstein2009,Eckstein2010b}, the electronic relaxation is accelerated by increasing interaction strength. In contrast to the purely electronic case, however, the decay is here modulated by coherent phonon-induced oscillations.

This observation motivates a direct comparison between the dynamics of \(\Delta n(t)\) and that of local observables. Panel (b) shows the phonon dynamics through the variance of the phonon displacement relative to its post-quench thermal value, \(\langle XX\rangle-\langle XX\rangle_{\mathrm{th}}\), again for \(U/v_\ast=2\). This quantity exhibits damped oscillations associated with the coherent phonon motion induced by the quench. The corresponding phonon relaxation rate is extracted from the exponential decay of the envelope, as indicated by the dashed lines. Here the final thermal state is defined by the temperature corresponding to the conserved total energy after the quench. Almost the same damping rates are found for other local observables shown in the Supplementary Material, including the kinetic energy [Eq.~\eqref{eqn:Ekin}] and the phonon density [Eq.~\eqref{eqn:phononDensity}] in Fig.~\ref{fig:FigSupp1}.
% These quantities also display damped coherent oscillations, with a frequency set by the renormalized phonon mode extracted from Fig.~\ref{fig:Fig2supp}(b), and approach their corresponding thermal values on a timescale comparable to the phonon damping time. The oscillation amplitude becomes very small by \(t\approx 40\) for all local observables, including the ones not shown like the electron-phonon correlation [Eq.~\eqref{eqn:elphoncorr}].

The comparison of the two timescales is presented in Fig.~\ref{fig:Fig1}(c), where the solid lines denote the electronic decay rates extracted from panel (a), and the dashed lines the phonon damping rates obtained from panel (b), for \(U/v_\ast=0,1,2\). Two regimes emerge. For \(g_f\lesssim 0.5\), the phonon decay rate is larger than the electronic one, so the phonon oscillations are damped before the momentum distribution has thermalized. In this sense, the long-time dynamics is electron dominated, since the electrons constitute the bottleneck. For \(g_f\gtrsim 0.5\), the trend is reversed: the electronic decay rate overtakes the phonon one, so that the electrons relax first while the phonons remain the slow sector. The late-time dynamics is then phonon dominated. The smooth change between these two behaviors defines the thermalization crossover.

The main new result is that this crossover survives at finite \(U\). While the \(U=0\) case reproduces the Holstein-model behavior of Ref.~\cite{Murakami2015}, the curves for \(U=1\) and \(U=2\) show that moderate Hubbard correlations do not qualitatively alter the crossover. Their effect is mainly quantitative: in the electron-dominated regime (\(g_f\lesssim 0.5\)) the electronic decay rate increases with \(U\), whereas in the phonon-dominated regime (\(g_f\gtrsim 0.5\)) it decreases slightly with \(U\). By contrast, the phonon decay rates remain comparatively weakly affected. Thus, also at finite $U$, the approach to thermalization is primarily determined by which sector relaxes more slowly, namely whether the bottleneck is electronic or phononic.

\subsection{\label{sec:ResultsPt2}Emergence of thermalization in Step-by-Step DMFT}

\begin{figure*}
    \centering
    \includegraphics[width=\linewidth]{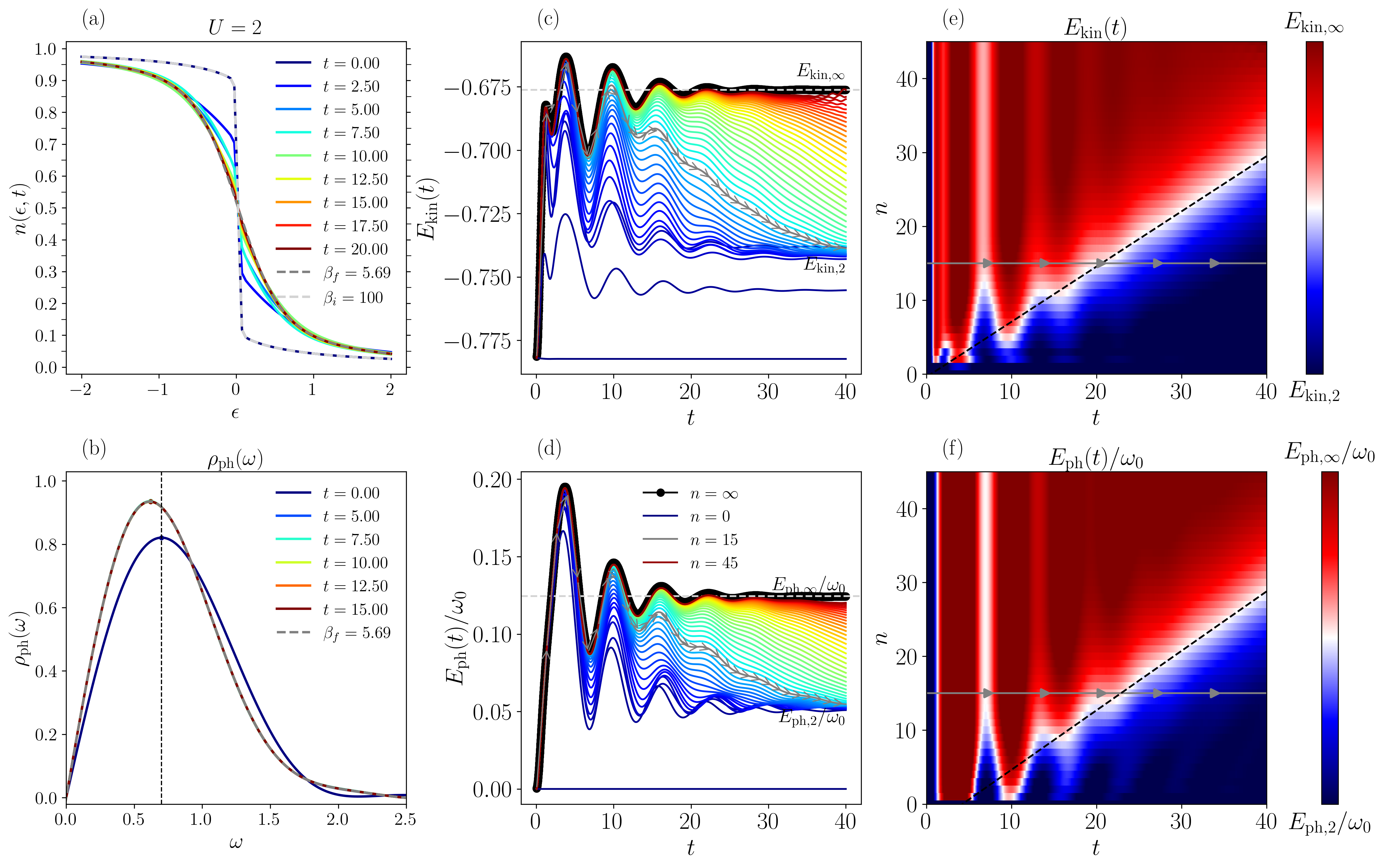}
     \caption{Thermalization in infinite dimensions for the Hubbard--Holstein model at $U/ v_\ast = 2$. 
Time evolution of the momentum distribution function $n(\epsilon,t)$ (a) and phonon spectral function $\rho_{\mathrm{ph}}(\omega)$ (b) after an interaction quench from $g=0$ to $g_f=0.5$ at $t=0^+$, starting from an initial thermal state at $\beta_i = 1/T_i =100$. 
Panels (c) and (d) show the time evolution of the kinetic energy $E_{\mathrm{kin}}(t)$ and free-phonon energy $E_{\mathrm{ph}}(t)/\omega_0$, respectively, for different DMFT iteration numbers (colored lines from blue to red as a function of the DMFT iteration number $n$), compared to the fully self-consistent DMFT solution ($n=\infty$, black line). 
Panels (e) and (f) display the emergence of a thermalization front in the kinetic and phonon energies as a function of time $t$ and DMFT iteration number $n$. 
The grey horizontal lines with arrows indicate a cut at fixed iteration number, highlighting the propagation of the thermalization front toward the fully self-consistent solution.
}
    \label{fig:Fig2}
\end{figure*}

We now analyze how thermalization emerges self-consistently within DMFT. We focus on the representative case \(U/v_\ast=2\) and \(g_f=0.5\), under the same conditions as above: half filling, \(\omega_0=0.7\), and initial inverse temperature \(\beta_i=100\). Results for other interaction strengths in the weak-coupling regime are discussed in the next section and in the Supplementary Material.

The fully converged nonequilibrium DMFT solution is shown in Fig.~\ref{fig:Fig2}(a,b) and in the black curves in panels (c,d) ($n=\infty$). Panel (a) displays the evolution of the momentum distribution \(n(\epsilon,t)\), which changes from the initial low-temperature form to a much smoother distribution approaching the thermal one at the final effective temperature \(\beta_f=1/T_{th}=5.69\) (dashed line). At the same time, the phonon spectral function,
\begin{equation}
\rho_{\mathrm{ph}}(\omega,t)=
-\frac{1}{\pi}\,\mathrm{Im}\int_{t-t_c}^{t}dt'\,
e^{i\omega(t-t')}\,
D^R(t,t'),
\end{equation}
with cutoff time \(t_c=10\), evolves toward the corresponding thermal spectrum [Fig.~\ref{fig:Fig2}(b)]. Starting from the bare phonon peak at \(\omega_0\) at \(t=0\), the spectrum develops a renormalized peak at a lower frequency \(\omega_0^r\), reflecting the dressing of the phonon mode by the electron-phonon interaction (Fig.~\ref{fig:Fig2supp})~\cite{Murakami2015}. The black curves in Fig.~\ref{fig:Fig2}(c,d) show the same time evolution of the kinetic energy and phonon density as in Fig.~\ref{fig:FigSupp1} for \(g_f=0.5\).

To understand how this thermal state emerges within DMFT, we adopt the Step-by-Step DMFT perspective introduced in Ref.~\cite{Picano2025_thermalization,Picano2025heating}. Instead of enforcing self-consistency at each time step, one first propagates the impurity dynamics in a fixed bath up to long times and only afterwards updates the bath for the next DMFT iteration \(n\). In this way, the long-time limit \(t\to\infty\) is effectively probed before full self-consistency \(n\to\infty\) is reached. This construction makes it possible to visualize how the lattice gradually acts as its own bath and how the final thermal state builds up through the DMFT loop.

The resulting dynamics is shown in Fig.~\ref{fig:Fig2}(c,d), where we plot the kinetic energy \(E_{\mathrm{kin}}(t)\) [Eq.~\eqref{eqn:Ekin}] and free-phonon energy \(E_{\mathrm{ph}}(t)/\omega_0\) [Eq.~\eqref{eqn:phononDensity}] for increasing DMFT iteration number \(n\), together with the fully converged solution (\(n=\infty\), black line).
For each finite \(n\), both observables initially follow the converged nonequilibrium DMFT trajectory, including the coherent oscillations associated with the dressed phonon mode. Once these oscillations are damped, however, the finite-\(n\) solutions eventually bend away from the converged curve and relax toward an intermediate long-time value, lying between the initial and final equilibrium energies. Thus, at fixed \(n\), the impurity does not remain at the final thermal state indefinitely, but returns at long times to the equilibrium state selected by the bath available at that iteration.

This behavior differs from the pure Hubbard case discussed in Ref.~\cite{Picano2025_thermalization}, where finite-\(n\) local observables relax back to the initial equilibrium value. Here the long-time plateau at finite \(n\) is instead shifted away from the initial one, because electrons and phonons exchange energy even before full DMFT self-consistency is reached. In this sense, the impurity already experiences a partially heated environment at finite iteration number. As \(n\) increases, the time interval over which the dynamics stays close to the fully converged solution becomes longer, showing that the thermalized portion of the time evolution progressively extends to later times.

This is seen explicitly in Fig.~\ref{fig:Fig2}(e,f), where the same observables are represented in the \((n,t)\) plane. After the short-time coherent oscillations in both the step-by-step kinetic energy and phonon density, one clearly observes a sharp crossover line, or thermalization front, separating two regimes: for \(t\lesssim \tau_n^\ast\), the dynamics is  close to the final thermal solution, while for \(t\gtrsim \tau_n^\ast\), it crosses over to the finite-\(n\) long-time behavior which remains close to the initial condition (horizontal grey line with arrows), but is shifted due to partial energy redistribution. The front propagates approximately linearly in \(n\), \(\tau_n^\ast\sim n/v\), indicating a ballistic spreading of thermalization through the DMFT iteration cycle. The same structure is visible in both the electronic and phononic energies, showing that the self-consistent buildup of thermalization involves the coupled electron-phonon system as a whole.
For \(g_f=0.5\), both quantities show almost  the same front velocity (slope of the dashed lines in panels (e,f)), indicating that thermalization propagates consistently through the coupled electron-phonon system. In the next section (and in Figs.~\ref{fig:FigSupp3}-~\ref{fig:FigSupp4} of the Supplemental Material), we investigate the dependence of this velocity on \(U\) and \(g_f\).

Figure~\ref{fig:Fig2} therefore provides a direct microscopic picture of how DMFT describes thermalization in the Hubbard-Holstein model: the impurity evolves in a bath that is progressively heated by the self-consistency condition, and the final thermal state emerges through a propagating front in the \((n,t)\) plane. In this way, the lattice acts as its own bath also in the presence of electron-phonon coupling, although the coexistence of electronic and phononic degrees of freedom modifies the finite-\(n\) long-time state compared with the purely electronic Hubbard case~\cite{Picano2025_thermalization}.

\subsection{\label{sec:ResultsPt3}Front velocity across the thermalization crossover}

\begin{figure*}
    \centering
    \includegraphics[width=\linewidth]{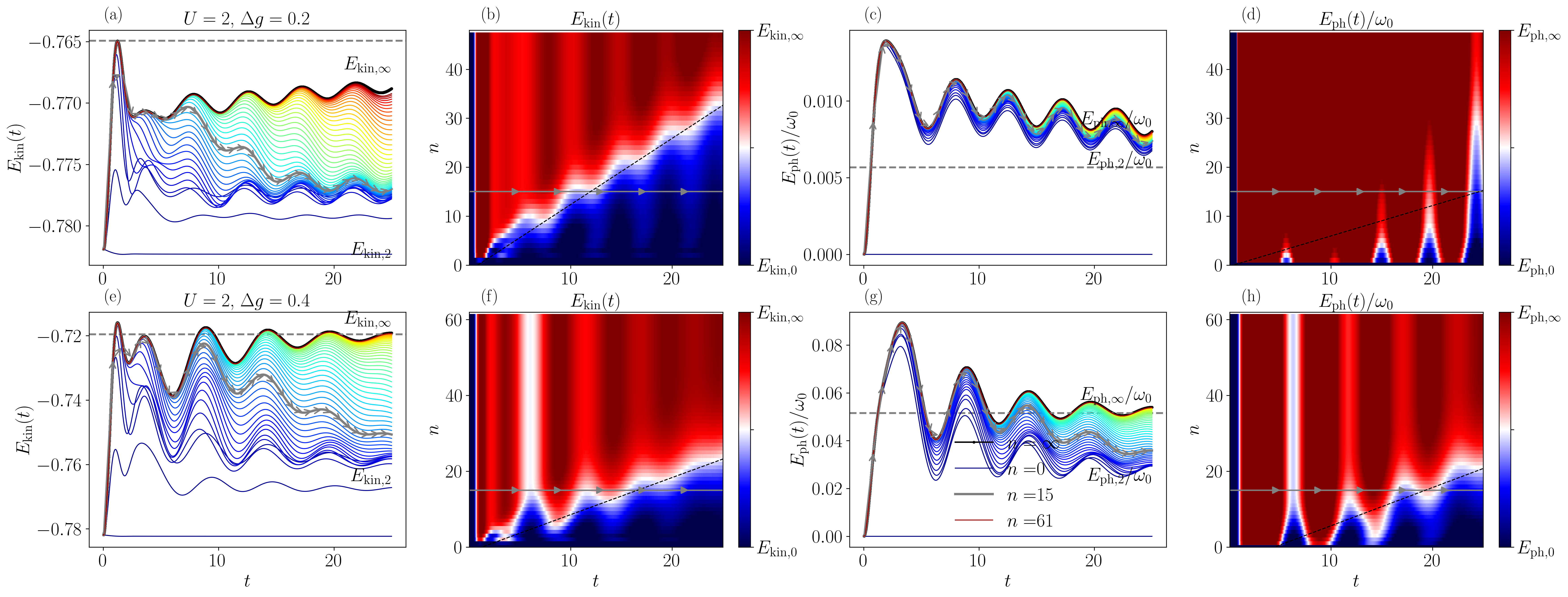}
    \caption{
Thermalization in infinite dimensions for the Hubbard--Holstein model at $U / v_\ast = 2$ for two different interaction quenches. 
Top row: weak quench from $g=0$ to $ g_f = 0.2$ at $t=0^+$. 
Bottom row: stronger quench from $g=0$ to $ g_f = 0.4$ at $t=0^+$. 
Panels (a,e) show the time evolution of the kinetic energy $E_{\mathrm{kin}}(t)$ for different DMFT iteration numbers (colored lines from blue to red as a function of the DMFT iteration number $n$), compared to the fully self-consistent DMFT solution ($n=\infty$, black line). 
Panels (c,g) display the corresponding free-phonon energy $E_{\mathrm{ph}}(t)/\omega_0$. 
Panels (b,f) and (d,h) show the emergence of a thermalization front in the kinetic and phonon energies, respectively, as a function of time $t$ and DMFT iteration number $n$. For the weak quench $g_f = 0.2$, no thermalization front develops in the free-phonon energy, as the step-by-step DMFT iterations rapidly collapse onto the fully self-consistent solution. 
The grey horizontal lines with arrows indicate cuts at fixed iteration number, while the dashed black lines highlight the propagation of the thermalization front toward the fully self-consistent solution.
}
    \label{fig:Fig3}
\end{figure*}

\begin{figure}
    \centering
    \includegraphics[width=\linewidth]{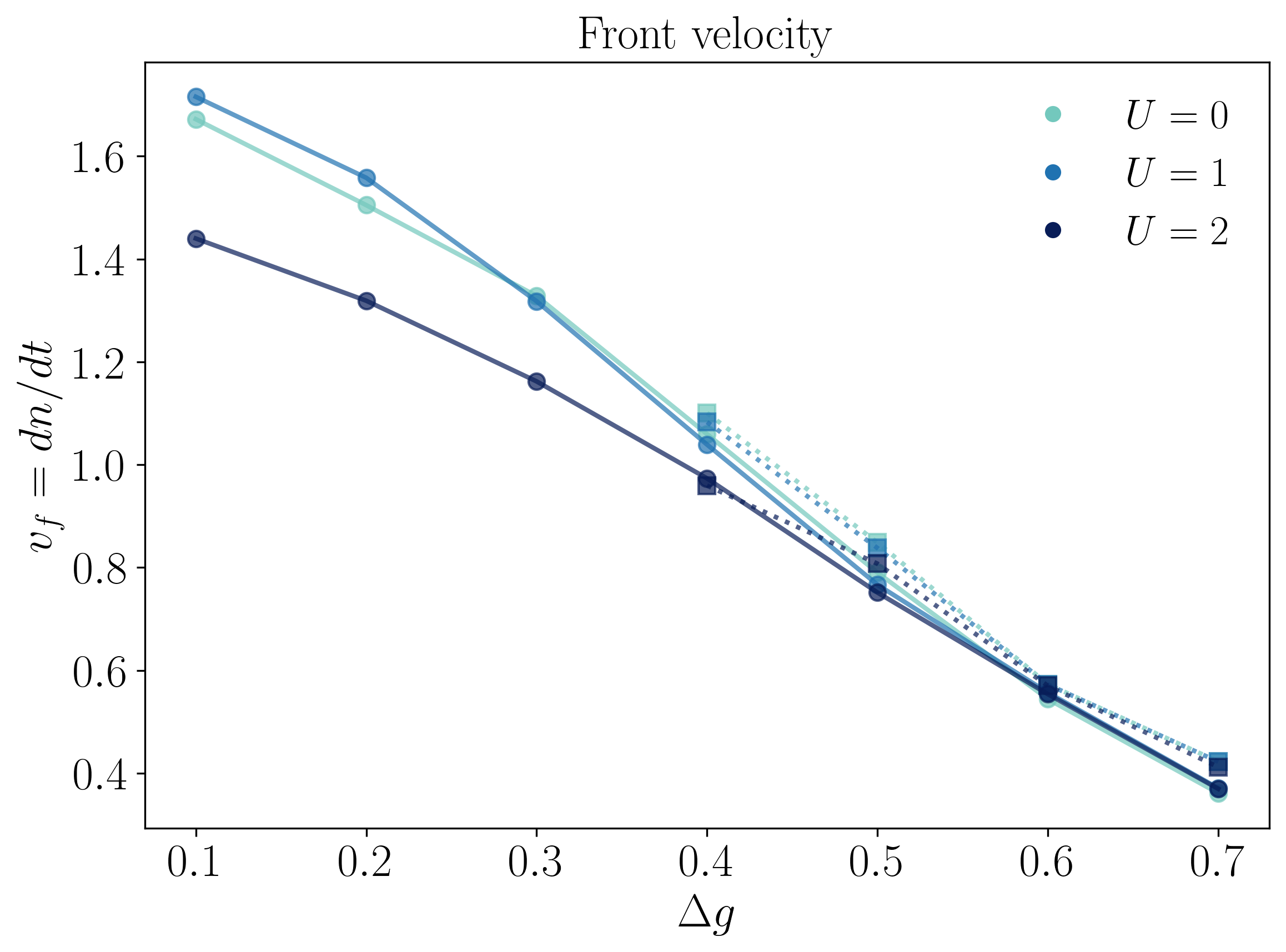}
   \caption{
Front velocity $v_f = dn/dt$ as a function of the interaction quench amplitude $\Delta g$ for different Hubbard interactions $U / v_\ast = 0, 1, 2$. 
Continuous lines denote the velocity extracted from the thermalization front in the electronic kinetic energy $E_{\mathrm{kin}}(t)$, while dotted lines indicate the velocity extracted from the free-phonon energy $E_{\mathrm{ph}}(t)$. 
For $\Delta g < 0.4$ no clear thermalization front develops in the phonon sector, and therefore the corresponding phonon velocities are not shown in that regime.
}
    \label{fig:Fig4}
\end{figure}

In Fig.~\ref{fig:Fig2}, the thermalization front in the kinetic energy and in the free-phonon energy appears after the same time offset and propagates with the same velocity. Figures~\ref{fig:Fig3} and \ref{fig:Fig4} show that this is not the case for all values of \(g_f\) and \(U\). For \(g_f=0.2\) [panels (a)--(d)], the electronic kinetic energy still displays a clear thermalization front in the \((n,t)\) plane [panel (b)], whereas no equally clear front is visible in the free-phonon energy within the accessible time window [panel (d)]. By contrast, for the stronger quench \(g_f=0.4\) [panels (e)--(h)], a front is clearly visible in both the electronic and phononic observables.

This difference reflects the asymmetric role of electrons and phonons in the DMFT self-consistency. The electronic bath, encoded in the hybridization function \(\Delta_n\), is updated directly through the electronic Green's function [Eq.~\eqref{eqn:Deltan}], so electronic observables naturally exhibit the step-by-step thermalization mechanism already for weak quenches. The phonons, by contrast, are local and dispersionless and are not coupled directly to the DMFT bath \(\Delta_n\). Their thermalization front can therefore emerge only indirectly, through the electron-phonon interaction via the phonon self-energy, i.e. the polarizability. For small \(g_f\), this effect is too weak to generate a clear phonon front on the simulated timescale [panels (c),(d)]. Only when \(g_f\) becomes sufficiently large, i.e. close to and beyond the thermalization crossover, does the phonon sector thermalize on a timescale comparable to the electronic one and develop a visible ballistic front [panels (g),(h)]. See also Fig.~\ref{fig:FigSupp4} of the Supplemental Material for the corresponding thermalization at \(U=0\).

This behavior is summarized in Fig.~\ref{fig:Fig4}, where we plot the front velocity \(v_f=dn/dt\) as a function of the quench amplitude \(\Delta g\) for \(U/v_\ast=0,1,2\). The velocity extracted from the electronic kinetic energy is shown for the full range of couplings, while the phonon velocity is reported only where a clear front can be identified. For \(\Delta g\gtrsim 0.4\), the electronic and phononic velocities almost coincide for all three values of \(U\). This shows that, once the electron-phonon coupling is sufficiently strong for the phonon sector to participate in the step-by-step thermalization process on comparable timescales, the thermalization front propagates coherently through the coupled system.
A second important trend in Fig.~\ref{fig:Fig4} is that the front velocity decreases as \(\Delta g\) increases. Since the front velocity measures how many DMFT iterations are needed to erase the memory of the initial condition at a given time, a smaller \(v_f\) corresponds to faster thermalization in the DMFT loop. This is consistent with the direct real-time dynamics discussed above: increasing \(g_f\) accelerates the decay of electronic observables, and correspondingly reduces the number of DMFT iterations required for the thermalized solution to emerge.

Finally, the dependence on \(U\) is comparatively weak. At fixed \(\Delta g\), increasing \(U\) slightly reduces the front velocity, but does not qualitatively modify the overall picture. The main control parameter for the propagation of the thermalization front is therefore the strength of the electron-phonon quench, while moderate Hubbard interactions only weakly renormalize the corresponding scale.

\section{\label{sec:Conclusion}Conclusion}

In this work, we studied the nonequilibrium dynamics of the weak-coupling Hubbard-Holstein model after a sudden switch-on of the electron-phonon interaction within nonequilibrium DMFT. The impurity problem was solved using the self-consistent Migdal approximation for the electron-phonon coupling, which captures the feedback of the electronic dynamics on the phonons through the phonon self-energy, together with second-order perturbation theory for the electron-electron interaction.
Extending the $U=0$
Holstein-model analysis to finite Hubbard interaction, we found that the relaxation dynamics is still controlled by the competition between electronic and phononic timescales. Comparing the decay of the jump 
$\Delta n (t)$ in the electronic momentum distribution with the damping of the phonon oscillations, we identified a thermalization crossover between an electron-dominated regime at weaker coupling and a phonon-dominated regime at stronger coupling. This crossover remains robust for 
$U/v_\ast = 0, 1,2 $ indicating that moderate local correlations do not qualitatively change the thermalization mechanism, but only weakly renormalize the associated decay rates.

The main result of this manuscript concerns the microscopic emergence of thermalization within DMFT in the weak-coupling Hubbard-Holstein model. Using the Step-by-Step DMFT construction introduced in Ref.~\cite{Picano2025_thermalization}, we showed that the thermal state builds up progressively through the self-consistency loop. In the \((n,t)\) plane, after the initial coherent oscillations of the local observables, this process is marked by a sharp thermalization front separating the region already described by the final thermal solution from the region that still retains memory of the initial condition. This front appears clearly in electronic observables already for weak quenches, whereas in the phononic sector it becomes visible within the accessible time window only once the electron-phonon coupling is sufficiently strong, i.e. close to and beyond the thermalization crossover. At weaker coupling, the local dispersionless Holstein phonons display a delayed onset of step-by-step thermalization, since they are driven only indirectly by the electronic sector. Whenever both fronts are visible, they propagate with the same velocity, indicating that thermalization spreads coherently through the coupled electron-phonon system.

More generally, our analysis shows that DMFT provides not only the final thermalized dynamics, but also a direct self-consistent picture of how thermalization emerges in an isolated lattice system acting as its own bath. In the present Hubbard-Holstein case, this mechanism is enriched by the interplay between electronic and phononic relaxation channels, which determines both the long-time dynamics and the propagation of the thermalization front.

Our work provides  a starting point for further studies of electron-phonon systems. 
The present analysis neglects the momentum dependence of the electron and phonon self-energies, an approximation  justified in the limit of large coordination.
In finite dimensions, however, momentum-dependent effects and short-range spatial correlations 
are expected to become important.
A natural extension would therefore be to move beyond single-site DMFT and consider cluster nonequilibrium methods, such as nonequilibrium dynamical cluster approximation (DCA)~\cite{Biroli2002,Aryanpour_2003}, in order to assess the role of nonlocal correlations and momentum-dependent phonon effects.
It would also be interesting to explore stronger couplings, the vicinity of the bipolaronic and Mott transitions~\cite{Eckstein2010}, and the nonequilibrium dynamics of symmetry-broken phases in coupled electron-phonon systems~\cite{Randi2017}.

Another natural direction concerns disordered electron-phonon systems, which could be studied within statistical~\cite{Miranda2011} or inhomogeneous~\cite{Picano2021_CDW,Picano2023_semicl,Valiera2025_SR} DMFT. In the presence of strong disorder, one may expect the ballistic thermalization front to be replaced by a localized one that retains memory of the initial condition. More generally, in the simultaneous presence of disorder and interactions, the competition between propagating thermalization fronts and disorder-induced pinning may provide a useful perspective on the stability of many-body localization.

\begin{acknowledgments}
M.S. acknowledges financial support
from the ERC consolidator Grant No. 101002955 - CONQUER. A. P. acknowledges funding from the
European Union’s Horizon 2020 research and innovation programme under the Marie Sklodowska-Curie Postdoctoral
Fellowship (Grant
Agreement No. 101149691 - DISRUPT).
A. P. acknowledges Jules Sueiro and Matthieu Vanhoecke for fruitful discussions.  
We are grateful for the use of computational resources from the Collège de France IPH cluster and  the Adastra supercomputer hosted at CINES. A. P. extends thanks to Philipp Hansmann and the RRZE of the University of Erlangen-Nuremberg for generously providing additional computational resources.
\end{acknowledgments}

\bibliography{sources}

\newpage
\appendix

%%%%%%%%%%%%%%%%%%%%%%%%%%%%%%%%
%\section{\label{sec:Appendix} Appendix} 

%%%%%%%%%%%%%%%%%%%%%%%%%%%%%%%%%%%%%%%%%%%%%%%%%%%%%%%%%%%%%%%%%%%%%%%%%%%%%%%%%%%%%%%%%%%%%%%

 \section{Local relaxation dynamics and equilibrium spectra after the quench}

\begin{figure*}
    \centering
    \includegraphics[width=\linewidth]{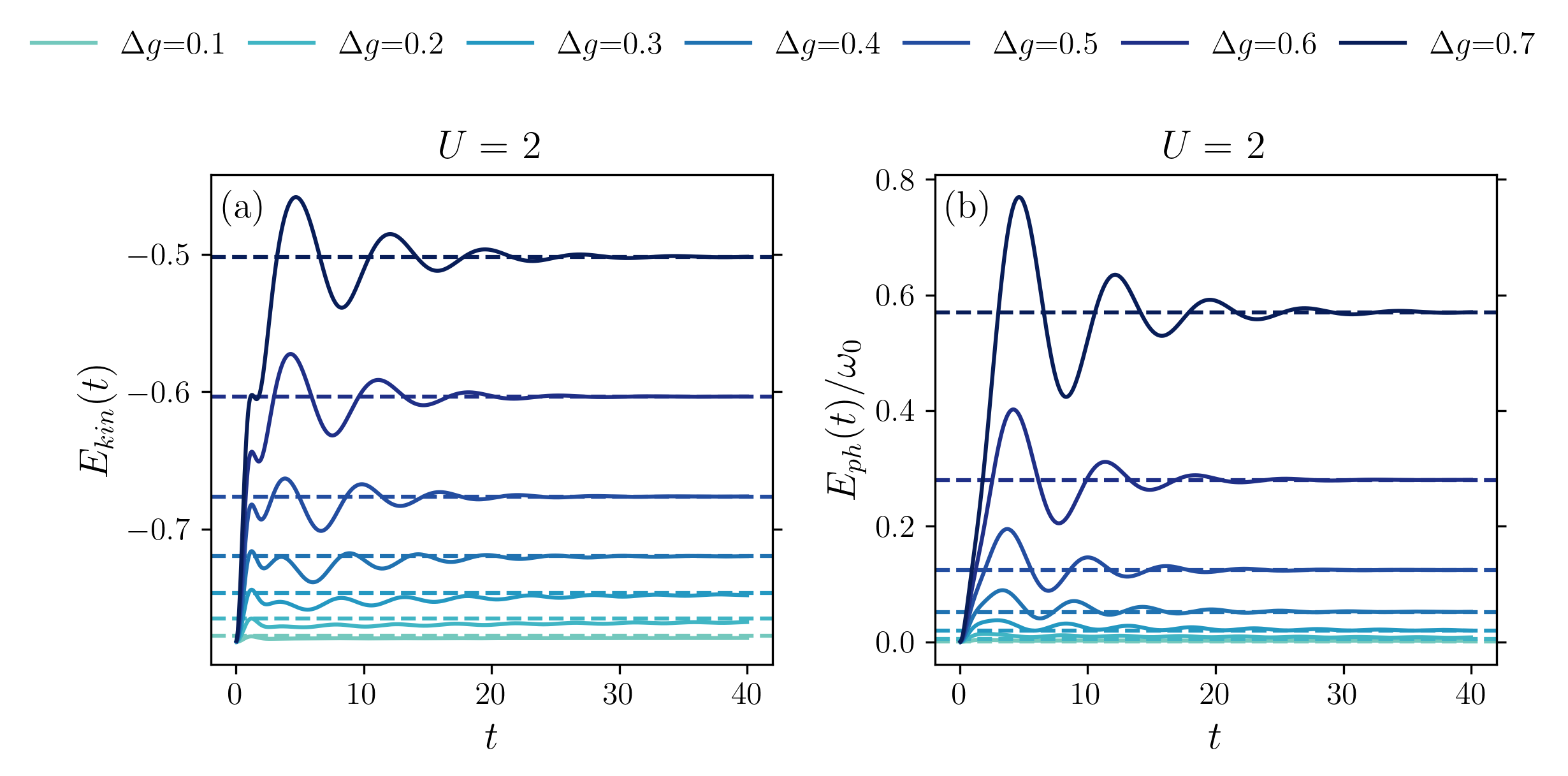}
    \caption{Time evolution of the kinetic energy $E_{\mathrm{kin}}(t)$ (left panel) and free-phonon energy $E_{\mathrm{ph}}(t)/\omega_0$ (right panel) after a quench in the electron-phonon coupling from $g=0$ to $g_f$, for $U=2$ and several values of the final coupling $g_f$. The dashed lines denote the corresponding thermal values for each $g_f$.
}
    \label{fig:FigSupp1}
\end{figure*}

\begin{figure*}
    \centering
    \includegraphics[width=\linewidth]{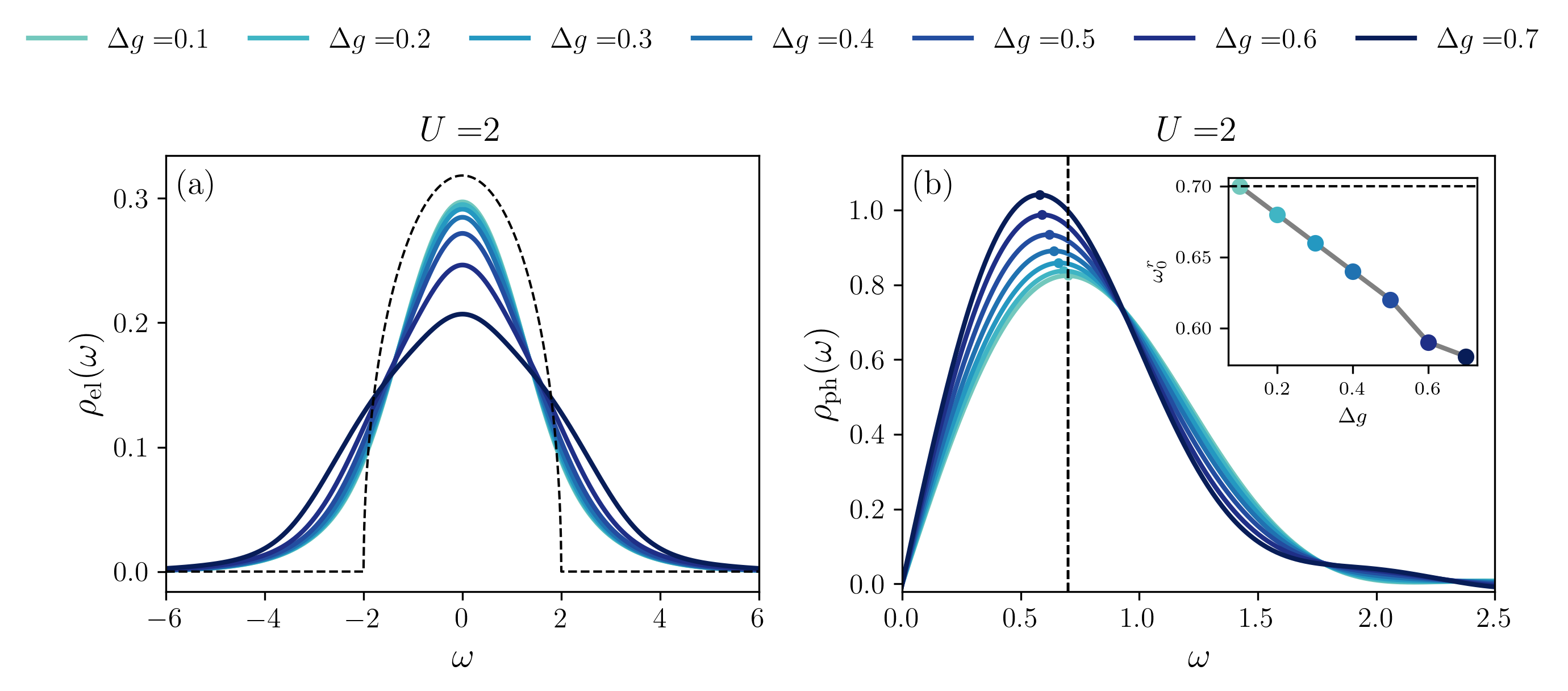}
\caption{The electron spectral functions $\rho_{\mathrm{el}}(\omega)$ [panel (a)] and phonon spectral functions $\rho_{\mathrm{ph}}(\omega)$ [panel (b)] for $U/v_\ast=2$, evaluated in equilibrium at the post-quench effective temperature $T_{\mathrm{th}}$ corresponding to the conserved total energy after a quench of the electron-phonon coupling from $g=0$ to $g_f$. Results are shown for several final coupling values $g_f$. The dashed black curves/lines indicate the corresponding noninteracting reference features: in the left panel, the bare electronic density of states, and in the right panel, the bare phonon frequency $\omega_0$. The inset in the right panel shows the position $\omega_0^r$ of the renormalized phonon peak as a function of $g_f$.}
    \label{fig:Fig2supp}
\end{figure*}

In Fig.~\ref{fig:FigSupp1}, we show the time evolution of the kinetic energy and the phonon density \(\langle a^\dagger a\rangle\) for \(U/v_\ast=2\). In analogy with the \(U=0\) case discussed in Ref.~\cite{Murakami2015}, both quantities exhibit coherent oscillations after the quench. In the present correlated case, these oscillations remain clearly visible and occur at a frequency set by the renormalized phonon mode \(\omega_0^r\), extracted from Fig.~\ref{fig:Fig2supp}.

As time evolves, the oscillations are damped and their amplitude becomes very small by \(t\approx 40\) for all values of \(g_f\) shown here. The dashed lines in Fig.~\ref{fig:FigSupp1} indicate the corresponding thermal expectation values at the post-quench temperature \(T_{\mathrm{th}}\), determined from the conserved total energy after the quench. Once the oscillations are sufficiently damped, the local observables are already very close to these thermal values. This shows that, also for finite \(U\), momentum-integrated quantities relax efficiently toward equilibrium. As discussed in the main text, however, this does not necessarily imply full thermalization, since momentum-resolved electronic observables may remain nonthermal on longer timescales.

Figure~\ref{fig:Fig2supp} provides the corresponding spectral information for \(U/v_\ast=2\). The electronic spectral function \(\rho_{\mathrm{el}}(\omega)\), evaluated at the effective temperature \(T_{\mathrm{th}}\) corresponding to the conserved post-quench energy, broadens and redistributes weight around the Fermi level as \(g_f\) increases, signaling the growing influence of the electron-phonon coupling on the low-energy electronic states. At the same time, the phonon spectral function \(\rho_{\mathrm{ph}}(\omega)\) shows a clear softening of the phonon peak relative to the bare frequency \(\omega_0\), as highlighted in the inset. The decrease of the renormalized phonon frequency \(\omega_0^r\) with increasing \(g_f\) is consistent with the oscillation timescales observed in Fig.~\ref{fig:FigSupp1} and supports the interpretation of the local dynamics in terms of coherent motion of the dressed phonon mode.

 \subsection{Step-by-Step DMFT at \(U=0\)}

\begin{figure*}
    \centering
    \includegraphics[width=\linewidth]{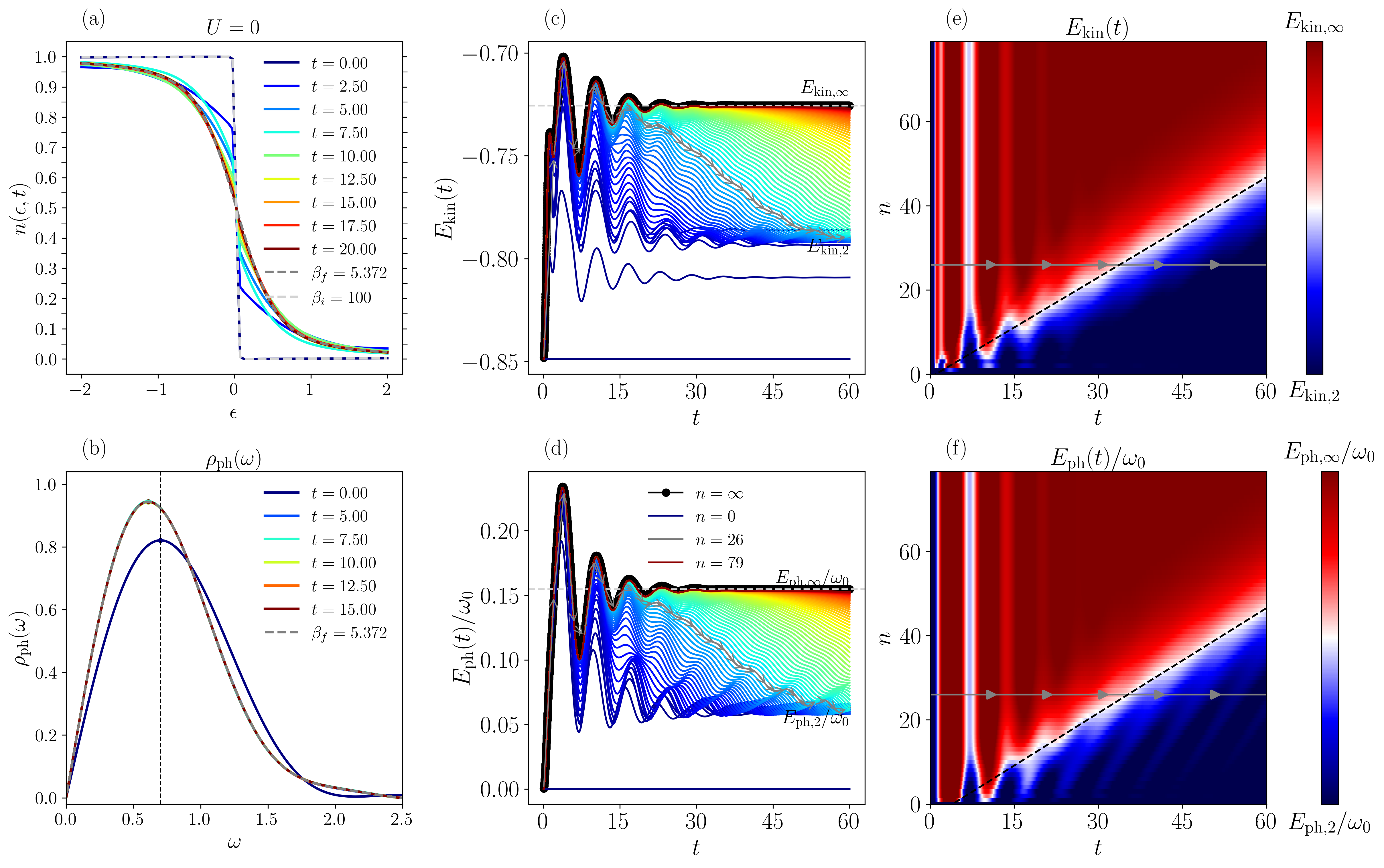}
\caption{
Thermalization in infinite dimensions for the Hubbard--Holstein model at $U/ v_\ast = 0$. 
Time evolution of the momentum distribution function $n(\epsilon,t)$ (a) and phonon spectral function $\rho_{\mathrm{ph}}(\omega)$ (b) after an interaction quench from $g=0$ to $g_f=0.5$ at $t=0^+$, starting from an initial thermal state at $\beta_i = 1/T_i =100$. 
Panels (c) and (d) show the time evolution of the kinetic energy $E_{\mathrm{kin}}(t)$ and free-phonon energy $E_{\mathrm{ph}}(t)/\omega_0$, respectively, for different DMFT iteration numbers (colored lines from blue to red as a function of the DMFT iteration number $n$), compared to the fully self-consistent DMFT solution ($n=\infty$, black line). 
Panels (e) and (f) display the emergence of a thermalization front in the kinetic and phonon energies as a function of time $t$ and DMFT iteration number $n$. 
The grey horizontal lines with arrows indicate a cut at fixed iteration number, highlighting the propagation of the thermalization front toward the fully self-consistent solution.
}
    \label{fig:FigSupp3}
\end{figure*}

\begin{figure*}
    \centering
    \includegraphics[width=\linewidth]{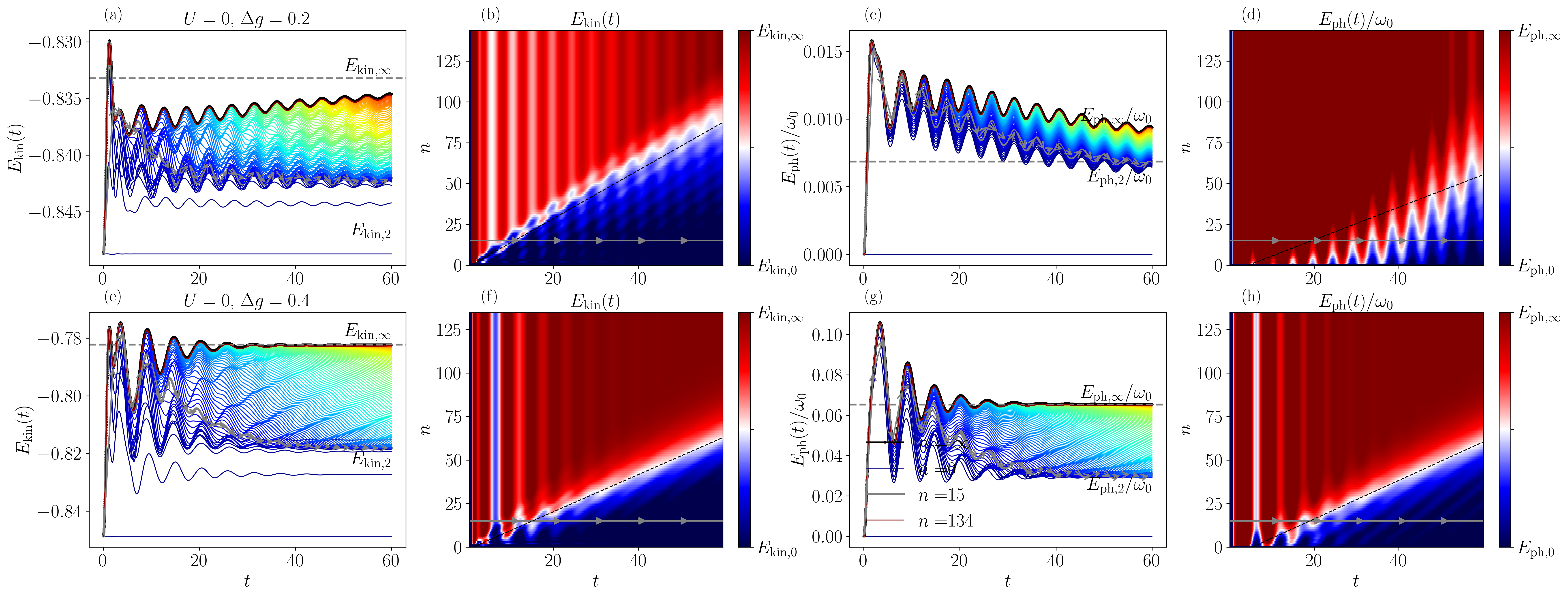}
    \caption{
Thermalization in infinite dimensions for the Hubbard--Holstein model at $U / v_\ast = 0$ for two different interaction quenches. 
Top row: weak quench to $g_f = 0.2$. 
Bottom row: stronger quench to $g_f = 0.4$. 
Panels (a,e) show the time evolution of the kinetic energy $E_{\mathrm{kin}}(t)$ for different DMFT iteration numbers (colored lines from blue to red as a function of the DMFT iteration number $n$), compared to the fully self-consistent DMFT solution ($n=\infty$, black line). 
Panels (c,g) display the corresponding free-phonon energy $E_{\mathrm{ph}}(t)/\omega_0$. 
Panels (b,f) and (d,h) show the emergence of a thermalization front in the kinetic and phonon energies, respectively, as a function of time $t$ and DMFT iteration number $n$. 
For the weak quench $g_f = 0.1$, a thermalization front in the free-phonon energy develops only at later times ($t \gtrsim 30$), after the initial oscillatory transient. 
The horizontal dashed green lines indicate the final equilibrium energy values at the effective temperature $T_f$ after the quench. 
The grey horizontal lines with arrows indicate cuts at fixed iteration number, while the dashed black lines highlight the propagation of the thermalization front toward the fully self-consistent solution.
}
    \label{fig:FigSupp4}
\end{figure*}

For completeness, we briefly discuss the corresponding Step-by-Step DMFT results in the Holstein limit \(U/v_\ast=0\), shown in Figs.~\ref{fig:FigSupp3} and \ref{fig:FigSupp4}. We keep the same parameters as in the main text, namely half filling, \(\omega_0=0.7\), and initial inverse temperature \(\beta_i=100\).

Figure~\ref{fig:FigSupp3} shows the representative case \(g_f=0.5\). Panels (a) and (b) confirm that the fully converged DMFT dynamics leads to thermalization also in the absence of Hubbard interaction: the electronic momentum distribution \(n(\epsilon,t)\) evolves from the initial low-temperature profile toward the thermal distribution at the final effective temperature, while the phonon spectral function develops a renormalized peak at a softened phonon frequency. Panels (c) and (d) display the step-by-step evolution of the kinetic and phonon energies for increasing DMFT iteration number \(n\). As in the interacting case, the finite-\(n\) solutions initially follow the fully self-consistent trajectory, including the coherent oscillations, and only at later times cross over to their finite-\(n\) long-time values. This behavior generates a clear thermalization front in the \((n,t)\) plane [panels (e) and (f)], which propagates approximately linearly in \(n\).

Figure~\ref{fig:FigSupp4} illustrates how this picture evolves with the quench strength. For the weaker quench, the thermalization front is clearly visible in the kinetic energy, while in the free-phonon energy it emerges only at later times, after the initial oscillatory transient. For the stronger quench, by contrast, the front is well developed in both observables and propagates with the same velocity. These results are fully consistent with the main text: even in the \(U=0\) limit, thermalization emerges within DMFT through a propagating front in the \((n,t)\) plane.
As the strength of the electron-phonon quench increases, the front propagates faster, so that the memory of the initial condition is erased on shorter timescales.

%\bibliography{sources.bib}

\end{document}